\documentclass[a4paper,aps,pra,showpacs,twocolumn,superscriptaddress]{revtex4-1}   
   
\usepackage[utf8]{inputenc}  
\usepackage[T1]{fontenc}     
\usepackage[british]{babel}  
\usepackage{lmodern}  
\usepackage[scaled=1.03]{inconsolata} 
\usepackage[usenames,dvipsnames]{color} 
\usepackage[colorlinks,citecolor=blue,linkcolor=magenta,urlcolor=blue]{hyperref}  
\usepackage{graphicx} 
\usepackage{tikz}
\usepackage[babel]{microtype}  
\usepackage{amsmath,amssymb,amsthm,bm,mathtools,amsfonts,mathrsfs,bbm,dsfont} 
\usepackage{xspace}  
\usepackage{multirow}
\usepackage{verbatim}

\usepackage{physics}
\newcommand{\id}{\ensuremath{\mathds{1}}}
\usepackage{bbold}

\newtheoremstyle{mystyle}
  {6pt}
  {6pt}
  {\normalfont}
  {0pt}
  {\bf}
  {.}
  { }
  {}

\theoremstyle{mystyle}
\newtheorem{theorem}{Theorem}

\begin{document}
\nonfrenchspacing
\title{Quantification of quantum dynamics with input-output games}

\author{Roope Uola}
\affiliation{D\'{e}partement de Physique Appliqu\'{e}e, Universit\'{e}  de Gen\`{e}ve, CH-1211 Gen\`{e}ve, Switzerland}

\author{Tristan Kraft}
\affiliation{Naturwissenschaftlich-Technische Fakult\"at, Universit\"at Siegen, Walter-Flex-Str. 3, D-57068 Siegen, Germany}

\author{Alastair A.\ Abbott}
\affiliation{D\'{e}partement de Physique Appliqu\'{e}e, Universit\'{e}  de Gen\`{e}ve, CH-1211 Gen\`{e}ve, Switzerland}

\date{\today}

\begin{abstract}
Recent developments surrounding resource theories have shown that any quantum state or measurement resource, with respect to a convex (and compact) set of resourceless objects, provides an advantage in a tailored subchannel or state discrimination task, respectively. Here we show that an analogous, more general result is also true in the case of dynamical quantum resources, i.e., channels and instruments. In the scenario we consider, the tasks associated to a resource are input-output games. The advantage a resource provides in these games is naturally quantified by a generalized robustness measure. We illustrate our approach by applying it to a broad collection of examples, including classical and measure-and-prepare channels, measurement and channel incompatibility, LOCC operations, and steering, as well as discussing its applicability to other resources in, e.g., quantum thermodynamics. We finish by showing that our approach generalizes to higher-order dynamics where it can be used, for example, to witness causal properties of supermaps.
\end{abstract}

\maketitle

\section{Introduction}

The advantage of quantum information processing over its classical counterpart has become evident over the previous decades. There are numerous tasks known for which a quantum resource is needed in order to gain an advantage over all classical protocols. For example, in quantum key distribution~\cite{Bennett1984, Ekert1991} entanglement is necessary for unconditionally secure key generation~\cite{Curty2004}, while it is also a resource for teleportation~\cite{Bennett1993} and measurement based quantum computation~\cite{Raussendorf2001} amongst many other tasks.

Whereas some quantum resources have been proven also to be sufficient for certain tasks, e.g., entanglement for randomness certification~\cite{Curchod2017}, and Bell-nonlocality for communication complexity protocols~\cite{Brukner2004}, no resource is expected to be useful for every task. This raises the question of which tasks require a given resource and leads to the notion of resource theories~\cite{Chitambar2019}. Resource theories are defined through free objects and free operations. Free objects are those that do not possess a given resource while free operations are transformations that leave the set of free objects invariant. As an example, in the resource theory of entanglement, the free objects are separable states while the free operations are local operations assisted by classical communication (LOCC)~\cite{Chitambar2014}.

Previously, much effort has been devoted to constructing resource theories for properties of quantum states, such as coherence~\cite{Aberg2006, Baumgratz2014, Streltsov2017}, reference frames \cite{bartlett2007, gour2008}, thermodynamical properties~\cite{Brandao2013}, and utility for stabilizer quantum computation~\cite{Veitch2014, Howard2017}. Here, we want to focus our attention on objects describing the dynamics of quantum systems, e.g., channels and instruments.

We develop an extremely general technique for finding tasks that certify dynamical quantum resources, and which encompasses non-dynamical resources that have previously been studied \cite{Aberg2006, Baumgratz2014, Streltsov2017,bartlett2007, gour2008,Brandao2013,Veitch2014, Howard2017,Piani2015,Carmeli2019,Skrzypczyk2019b,Oszmaniec2019,Guerini2019,Takagi2019,uola2019} such as states and measurements as special cases, thereby providing a unifying framework to certify such resources. 
More precisely, we show that dynamical quantum resources can outperform their corresponding resourceless objects in tailored input-output games. In these games, one party inputs a state from an ensemble into a channel, another party performs a measurement on the output, and different input-output pairs are given a score. The framework induces a natural quantifier for this outperformance---the generalized robustness---and we will show how this quantifier relates to the highest obtainable payoff in input-output games. 

\begin{figure}[b!]
    \centering
    \includegraphics[scale=.47]{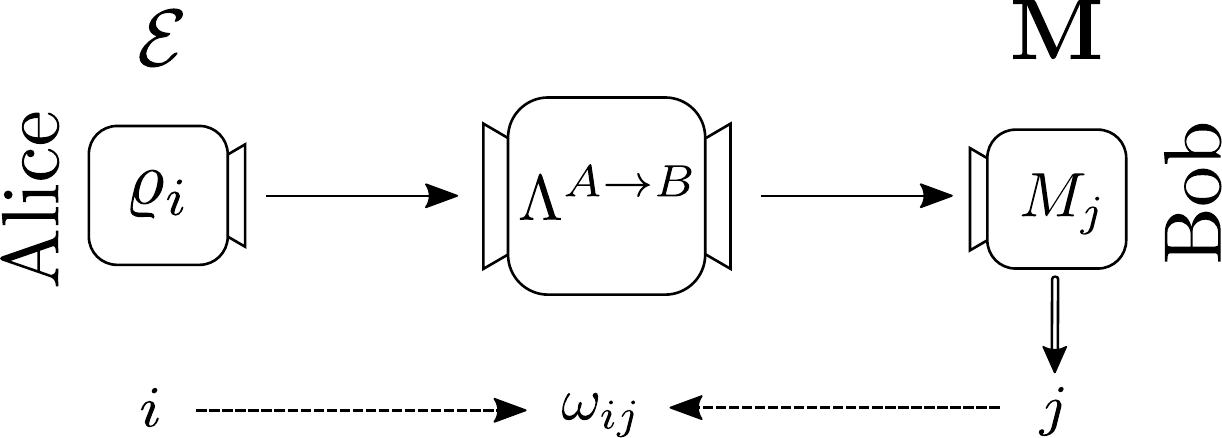}
    \caption{A simple input-output game: Alice inputs a state from an \textit{a priori} known ensemble $\mathcal{E}=\qty{p(i),\varrho_i}$ in to the channel $\Lambda^{A\rightarrow B}$. Bob performs a measurement $\textbf{M}=\qty{M_j}_j$ on the output of the channel. The goal is to choose the $\Lambda^{A\rightarrow B}$ maximizing the overall payoff $P(\Lambda^{A\rightarrow B}, \mathcal{E},\textbf{M},\Omega)$ that depends on the input ensemble, the channel, the measurement performed on the output and the reward function $\Omega=\qty{\omega_{ij}}_{ij}$.}
    \label{Fig:inout}
\end{figure}

We exhibit the generality of our approach by applying it to several examples, including properties of quantum channels related to breaking of entanglement~\cite{Horodecki2003, Simnacher2019} as well as incompatibility~\cite{Heinosaari2015}.
Beyond properties of individual dynamical objects, our technique is also applicable to sets of channels, quantum instruments, and to higher-order dynamics, (e.g., supermaps and superinstruments~\cite{chiribella08}), objects for which such operational advantages have not previously been identified. This results in simple, operationally motivated quantifiers for resources such as incompatibility of channels and testers~\cite{ziman08}, maps unreachable by local operations assisted by classical communication, and causal nonseparability~\cite{oreshkov12}.

\section{Input-output games}

The operational tasks we will use to quantify quantum dynamical resources are generalizations of quantum discrimination games, called input-output games.
Consider two players Alice and Bob, both of which receive an input label $x$ in each round of the game. Upon receiving $x$ Alice randomly prepares a state from the state assemblages $\mathbf{A}=\{p(i,x),\varrho_{i|x}\}_{i,x}$ and sends her state though a channel from the collection $\mathbf{\Lambda}=\qty{\Lambda_x^{A\rightarrow B}}_{x=1}^{|X|}$ of $|X|$ channels to Bob. After receiving the output-state, Bob performs a measurement $x$ from the set of POVMs $\mathbf{M}=\{M_{j|x}\}_{j,x}$. For each input $x$, and pair of preparation $i$ and measurement result $j$ the players receive a score according to a reward function $\Omega=\{\omega_{ijx}\}_{i,j,x}$, where $\omega_{ijx}$ are real numbers.
The tuple $\mathcal{G}=(\mathbf{A},\mathbf{M}, \Omega)$ defines an input-output game, while the collection of channels $\mathbf{\Lambda}$ chosen by the parties is thus the ``strategy'' for the game. The quantifier of success then takes the form 
\begin{align}\label{eq:payoffDef}
P(\mathbf{\Lambda},\mathcal{G})=\sum_{i,j,x}p(i,x)\omega_{ijx}\tr[\Lambda_x^{A\rightarrow B}(\varrho_{i|x}) M_{j|x}].
\end{align}
For the case of a single channel, i.e. $|X|=1$, the game is illustrated in Fig.~\ref{Fig:inout}.

Any input-output game gives rise to a class of equivalent games obtained by scaling and shifting the payoff. 
In order to use such games to quantify dynamical resource it will be necessary to restrict ourselves to a class of ``canonical'' input-output games, for which $\min_\mathbf{\Lambda}P(\mathbf{\Lambda},\mathcal{G})=0$ and $\max_\mathbf{\Lambda}P(\mathbf{\Lambda},\mathcal{G})=1$. 
Note that any input-output game can be brought to this form, and henceforth 
we will implicitly assume that all input-output games are canonical.

It is interesting to note that for games with non-negative reward functions,
summing over $i$ in Eq.~\eqref{eq:payoffDef} gives a minimum-error discrimination task. Namely, defining the operators $\sigma_{j|x}:=\sum_i p(i,x)\omega_{ijx}\Lambda_x^{A\rightarrow B}(\varrho_{i|x})$ and a probability distribution $p(j,x):=\tr[\sigma_{j|x}]/N$ with $N=\sum_{j,x}\tr[\sigma_{j|x}]$ allows one to write $P(\mathbf{\Lambda},\mathcal{G})=N\sum_{j,x}p(j,x)\tr[\hat{\sigma}_{j|x}M_{j|x}]$, where $\hat{\sigma}_{j|x}$ is normalized. This is a minimum-error discrimination task with pre-measurement information, i.e., the information about the ensemble $x$ is known before choosing the measurement on the output.

\section{Quantification of resources}

In this section we introduce our main resource theoretical tool: the generalized robustness. The generalized robustness is a quantifier that measures the relative distance of an object from a convex and compact set of objects, called the free objects. Intuitively, it is the amount of noise needed to corrupt a resource. More precisely, we denote a convex and compact subset of collections of channels by $F$ (which includes channels as trivial collections) and call this the free set. The (generalized) robustness $\mathcal{R}_F(\mathbf{\Lambda})$ of a collection $\mathbf{\Lambda}$ with respect to the free set $F$ is defined as
\begin{equation}
\label{Eq:ChannelRobustness}
\mathcal{R}_F(\mathbf{\Lambda}) = \min_{\tilde{\mathbf{\Lambda}}}\qty{t\geq 0\,\bigg\vert\, \frac{\mathbf{\Lambda} + t\tilde{\mathbf{\Lambda}}}{1+t}\in F}.
\end{equation}
The optimization is over all collections $\tilde{\mathbf{\Lambda}}:=\qty{\tilde{\Lambda}_x^{A\rightarrow B}}_{x=1}^{|X|}$. By solving for $\Lambda$ in the above equation and using the linearity and positivity of the (canonical) payoff function one can write $P(\mathbf{\Lambda},\mathcal{G})=[1+\mathcal{R}_F(\mathbf{\Lambda})]P(\mathbf{\Gamma},\mathcal{G})-\mathcal{R}_F(\mathbf{\Lambda})P(\tilde{\mathbf{\Lambda}},\mathcal{G})\leq[1+\mathcal{R}_F(\mathbf{\Lambda})]P(\mathbf{\Gamma},\mathcal{G})$, where $\mathbf{\Gamma}\in F$. Hence, we arrive at
\begin{equation}
\label{Eq:PsuccRatio}
\frac{P(\mathbf{\Lambda},\mathcal{G})}{\max_{\mathbf{\Gamma}\in F} P(\mathbf{\Gamma}
,\mathcal{G})} \leq 1+\mathcal{R}_F(\mathbf{\Lambda}),
\end{equation}
where the maximization is taken over all free collections $\mathbf{\Gamma}\in F$.

Using the celebrated Choi isomorphism we can map any channel $\Lambda$ to a bipartite state $J_\Lambda = \frac{1}{d}\sum_{ij} \ketbra{i}{j}\otimes\Lambda[\ketbra{i}{j}]$ with a fixed marginal~\cite{Choi1975, Heinosaari2011}, where $d$ is the dimension of the channel input. As this mapping is one-to-one, one can evaluate the robustness within the image of the isomorphism, i.e., on a subset of bipartite quantum states. Using techniques developed in Refs.~\cite{uola2019, Takagi2019} (see Appendix~\ref{appendix:conic}) the robustness can be cast as a conic optimization problem
\begin{eqnarray}
\label{RobSetsChan}
1+\mathcal{R}_F(\mathbf{\Lambda}) = \max_{Y}\, &&  \sum_x \tr[Y_x J_{\Lambda_x^{A\rightarrow B}}] \\
\text{s.t.: }&& Y \geq 0, \quad \tr[YT]\leq 1\, \forall\ T\in J_{F}, \notag
\end{eqnarray}
where $Y=\oplus_x Y_x$ constitutes a witness, $J_{\Lambda_x^{A\rightarrow B}}$ denotes the Choi states of the channel $\Lambda_x^{A\rightarrow B}$, and $J_{F}$ is the image of the free set $F$ under the Choi isomorphism. Note that, in order to evaluate the robustness in the above form, there is a crucial assumption of Slater's condition being satisfied (see Appendix~\ref{appendix:conic} for details). We will implicitly assume that this holds---as is indeed the case in all the applications we consider---throughout the rest of this paper.

\section{Main result}

We have already seen that the generalized robustness measure and the advantage over the free set in input-output games are linked, cf.\ Eq.~(\ref{Eq:PsuccRatio}). In order to operationalize this link, one wishes to implement a witness as in Eq.~(\ref{RobSetsChan}) in a way that resembles an input-output game, cf.\ Eq.~(\ref{eq:payoffDef}). To do so, we write
\begin{equation}\label{eqn:Yx}
Y_x=d\sum_{ij}p(i,x)\omega_{ijx}\varrho_{i|x}^T\otimes \eta_{j|x}, 
\end{equation}
where $\varrho_{i|x}$ are quantum states, $\eta_{j|x}$ are positive semidefinite operators satisfying $\sum_{j=1}^n \eta_{j|x}\le \id$ for all $x$, $p(i,x)$ is a probability distribution and $\omega_{ijx}$ are real numbers~\cite{rosset2018}. Note that for every $x$ the collection $\{\eta_{j|x}\}_{j=1}^n$ can be completed into a POVM by adding an element $\eta_{n+1|x}:=\openone-\sum_{j=1}^n\eta_{j|x}$ for which the reward function is taken to be 0.
Hence, an optimal witness corresponds to an input-output game up to normalization. To see that the minimum value of the game is zero, one can solve $\Lambda$ from Eq.~(\ref{Eq:ChannelRobustness}) resulting in $\Lambda=[1+\mathcal{R}_F(\mathbf{\Lambda})]\Gamma-\mathcal{R}_F(\mathbf{\Lambda})\tilde\Lambda$ where $\Gamma\in F$. Putting the expression to the Choi picture, multiplying the resulting equation by an optimal witness and taking the trace on both sides gives in $\tr[Y J_{\tilde\Lambda}]=0$.
Noting that the normalization of a game does not affect Eq.~(\ref{Eq:PsuccRatio}), we combine Eq.~(\ref{Eq:PsuccRatio}) with Eq.~(\ref{RobSetsChan}) and write
\begin{equation}
\label{Eq:result1}
\sup_{\mathcal{G}}\frac{P(\mathbf{\Lambda},\mathcal{G})}{\max_{\mathbf{\Gamma}\in F} P(\mathbf{\Gamma},\mathcal{G})} = 1+\mathcal{R}_F(\mathbf{\Lambda}),
\end{equation}
where the supremum is taken over all input-output games $\mathcal{G}$. We have thus proven our main result:
\begin{theorem}\label{thm1}
	Let $F$ be a convex and compact set of collections of channels. For any $\mathbf{\Lambda}$ not in $F$ there exists a tailored input-output game $\mathcal{G}$ for which $\mathbf{\Lambda}$ outperforms any point in $F$. Moreover, this outperformance is quantified by the generalized robustness according to Eq.~\eqref{Eq:result1}.
\end{theorem}

An alternative approach to giving an operational characterization of single-channel resources was recently given in~\cite{Takagi2019a} by applying known results for state-resources to the Choi state of a channel.
Indeed, it is known that any state resource provides an advantage in a tailored subchannel discrimination task~\cite{Takagi2019} and in the Choi picture this shows that any channel resource can provide a similar advantage~\cite{Takagi2019a}.
Although similar in spirit, our approach presents two significant advantages as a suitable extension of the former results regarding state resources to the realm of dynamical objects.
Firstly, as we will see in Sec.~\ref{sec:supermaps}, our approach generalizes readily to higher-order dynamical objects and sets thereof, thereby providing a unified characterization of a wide array of different classes of resources. 
The game-theoretic characterization of these more general scenarios has not previously been considered in the literature, and such scenarios cannot be reduced to discrimination tasks in the same manner as channel resources.
Secondly, in contrast to the approach of~\cite{Takagi2019a}, our approach does not require applying channels on subsystems of larger, entangled states and performing joint measurements on such states.
The experimental resources required to experimentally implement the protocol are thus significantly reduced.

Finally, before discussing applications of our main result, let us first outline more explicitly how the tailored input-output game for a given resource can be constructed.
Firstly, one must obtain the optimal witness $Y$. In scenarios where the free set is characterised by positive semidefinite constraints, as for compatible channels or unsteerable instruments, it can be obtained by simply evaluating the robustness cone program with semidefinite programming techniques. This can be done efficiently using numerical methods. In scenarios where the free set is not characterised by such constraints, e.g.\ for incompatibility breaking channels, one can find a witness by heuristic methods such as semidefinite programming hierarchies. Then, in order to get the canonical input-output game from the obtained witness, one writes down the operator Schmidt decomposition for the witness together with the appropriate normalisation to put it in the form of $Y_x$ as in Eq.~\eqref{eqn:Yx}, thereby obtaining the elements of the game.

\section{Examples of channel resources}

In this section, we present task-oriented characterisations of resources related to quantum channels and sets thereof using our technique.

\subsection{Entanglement and incompatibility breaking channels}%
In the study of quantum correlations, one typically asks if a given quantum state can violate a classical criterion such as separability, unsteerability or a Bell inequality. Answering the converse question of whether a state belongs to some of these classes is typically very hard. However, alternative ways of characterizing the states satisfying the first two criteria are known and they relate naturally to our framework \cite{Piani2009,Piani2015}. 
As a first application of Theorem~\ref{thm1}, we focus on properties of channels related to separability and unsteerability.
The channels corresponding to states with these properties (through the inverse Choi isomorphism) are those that break the entanglement of all states (separability) or incompatibility of all measurements (unsteerability). Entanglement breaking channels are also known to coincide with measure-and-prepare channels~\cite{Horodecki2003}, whereas incompatibility breaking channels are so far lacking a simple characterization~\cite{Heinosaari2015}.

Both entanglement and incompatibility breaking channels form convex and compact subsets of channels and, hence, using our framework one can define the corresponding robustnesses and deduce a task-oriented characterization of these sets. We note that for entanglement breaking channels our result complements the witnessing techniques presented in Ref.~\cite{rosset2018}, where the authors develop a resource theory of quantum memories, i.e., channels that are not of the measure-and-prepare form, and discuss the implementation of measurement-device-independent witnesses for such channels. On top of a witness, our result provide a simple task-oriented quantifier for such memories. 
Our result can also be used to characterize interesting subsets of measure-and-prepare channels, such as those corresponding to POVMs, i.e., ones sending only a classical message. Formally, these channels can be written $\Lambda^{A\rightarrow B}(\varrho)=\sum_a\tr[N_a\varrho]|a\rangle\langle a|$, where $\qty{\ket{a}}$ is an orthonormal basis. Interestingly, this complements recent studies on semi-quantum games~\cite{Guerini2019} and measure-and-prepare scenarios by providing an alternative operational quantifier for the advantage a channel with a quantum message provides over all classical messages in a specific input-output game.

\subsection{Compatibility of channels}

A natural property of a set of channels is that of compatibility, i.e., the question whether a set of channels can be seen as part of a single channel. More precisely, a set of channels $\qty{\Lambda_{x}}_{x}$ is called compatible if there exists a broadcast channel $\Lambda$ such that $\Lambda_x=\tr_{\backslash x}[\Lambda]$~\cite{Heinosaari2015}. Clearly the set of compatible channels is convex, hence, fitting to the realm of Theorem~\ref{thm1}. 

As for entanglement breaking channels, an interesting special case is obtained when considering compatible sets of channels with trivial outputs. The compatibility of measure-and-prepare channels with trivial outputs corresponds to the compatibility of POVMs. Motivated by recent developments on the connection between compatibility of measurements and communication tasks~\cite{uola2019, Carmeli2019, Skrzypczyk2019b, Oszmaniec2019, Tavakoli2019, Guerini2019}, we spell out explicitly this example. A set $\qty{A_{a|x}}_{a,x}$ of POVMs is called compatible, or jointly measurable, if there exists a joint measurement $\qty{G_\lambda}_\lambda$ and probability distributions $p(a|x,\lambda)$ such that $A_{a|x}=\sum_\lambda p(a|x,\lambda)G_\lambda$. A set $\qty{A_{a|x}}_{a,x}$ of POVMs can be seen as a set of measure-and-prepare channels $\{\Lambda_x\}_x$ by defining $\Lambda_x^{A\rightarrow B}(\varrho)=\sum_a\tr[A_{a|x}\varrho]|a\rangle\langle a|$. The common channels are characterized as those that first measure a single POVM $\qty{G_\lambda}_\lambda$, produce a classical output $\lambda$ and post-process the output according to some probability distribution $p(a|x,\lambda)$. This indeed gives a one-to-one correspondence between compatible sets of POVMs and compatible sets of trivial-output channels \cite{Heinosaari2017}. In this way, joint measurability can be witnessed through input-output games. In the case of trivial output channels the witness formula takes a simpler form: 
\begin{align}
\sum_x \tr[Y_x J_{\Lambda_x}]&=\sum_{a,i,j,x}\omega_{ijx}\tr[A_{a|x}\varrho_{i|x}]\langle a|\eta_{j|x}|a\rangle\nonumber\\ 
&=\sum_{a,x}\tilde\omega_{ax}\tr[A_{a|x}\tilde\varrho_x],
\end{align}
where $\tilde\omega_{ax}\tilde\varrho_x:=\sum_{i,j}\omega_{ijx}\langle a|\eta_{j|x}|a\rangle\,\varrho_{i|x}$. One can further normalize the operators $\tilde\varrho_x$ by pushing the relevant factors into the payoff function. In this way, the input-output game becomes a witness of the incompatibility of the measurements. This shows that in the formalism of input-output games, incompatible measurements can perform better than compatible ones in measure-and-prepare scenarios where only classical information is sent forward, c.f.\ Ref.~\cite{Guerini2019} for a more detailed discussion on the connection between incompatibility and the quantumness of the sent message. Note that in the case of joint measurability, the explicit form of an optimal witness can be calculated via semidefinite programming.

\subsection{Further examples in other domains}
Besides the cases discussed in the previous section our methods can also be applied to other scenarios that have not yet been studied in the literature. The first example are so-called $G$-covariant operations.

Any transformation of a physical system requires a reference frame. For instance, a rotation of a qubit state on the Bloch sphere requires a notion of direction, i.e., asymmetry. On the contrary, lack of symmetry in the reference frame puts a restriction on what transformations can be implemented. Mathematically, the lack of symmetry is described by symmetry transformations~\cite{gour2008, bartlett2007}. Denote by $G$ the group of transformations that leaves the reference frame invariant and let $\mathcal{U}_g(\varrho)=U_g\varrho U_g^{\dagger}$ with $g\in G$ be a unitary representation of the group $G$. The \emph{$G$-covariant} operations $\Lambda$ that can be implemented under this restriction are those that commute with all symmetry transformations, i.e., $\qty[\Lambda, \mathcal{U}_g] = 0$ for all $g\in G$. The set of all $G$-invariant operations is convex and compact and hence the asymmetry of a channel can be witnessed using the approach the developed.

Another relevant example can be found in the context of quantum thermodynamics, namely that of decomposability into thermal operations. In quantum thermodynamics, thermal operations refer to a set of transformations that can be implemented without the need of an external source of work~\cite{Lostaglio2018}. Thermal operations are defined by $\mathcal{E}(\varrho) = \tr_R[U_{SR}(\varrho\otimes\tau_{\beta})U_{SR}^{\dagger}]$.
The initial state of the system $S$ is denoted by $\varrho$ and $\tau_{\beta}=\exp(-\beta H_R)/\tr[\exp(-\beta H_R)]$ is a Gibbs state of the reservoir. The global unitary transformation is such that $[U_{SR},H_S\otimes H_R]=0$, i.e., it is energy-preserving.
Theorem~\ref{thm1} shows that relevant classes of channels---for example, those that cannot be implemented as a convex mixture of sequences of thermal operations acting on lower-dimensional systems~\cite{Lostaglio2018,mazurek18}---can be harnessed to provide operational advantages in input-output games.

A third example is true quantum decoherence. Quantum decoherence can sometimes be explained by classical fluctuations in the ambient fields, i.e., by random unitary dynamics. However, in systems of dimension three or higher there exist decohering channels, i.e., unital channels, that are not of this form~\cite{landau93}. Such decoherence is sometimes referred to as one of true quantum nature \cite{Kayser2015}. Random unitary channels form a convex subset of channels and, hence, one can define a measure of true quantum decoherence (of a unital channel) as the generalized robustness with respect to random unitary channels. As with the previous examples, the possibility of true quantum decoherence can be witnessed through input-output games.

\section{Quantum instruments}

Before discussing the generalisation of our technique to higher-order dynamics, we explicitly formulate it for quantum instruments $\textbf{I}=\qty{I_{a|x}}_{a,x}$, i.e., collections of completely positive maps summing up to a channel. These are another crucial resource in quantum information that has not previously been given a general, task-based characterization. We define the robustness analogously to that in Eq.~\eqref{Eq:ChannelRobustness}. As in the case of channels, the robustness is preserved under the Choi isomorphism.

To make a connection between the robustness and input-output games, one writes the payoff function as $P(\textbf{I}, \mathcal{G})=\sum_{i,j,x,a} p(i,x,a)\omega_{ijxa}\tr[I_{a|x}(\varrho_{i|x,a}) M_{j|x,a}]$ and notices that a witness has the structure $Y=\oplus_{a,x}Y_{a|x}$. Note that every element $Y_{a|x}$ can be decomposed as $Y_{a|x}=d\sum_{i,j}p(i,x,a)\omega_{ijxa}\varrho_{i|x,a}^T\otimes \eta_{j|x,a}$.
In the Appendix~\ref{appendix:conic} we show that our Theorem \ref{thm1} holds true when replacing collections of channels with collections of instruments, thereby providing a way to witness resources based on quantum instruments using input-output games.

Note that in the case of instruments our input-output game is post-selected on the output $a$ of the instrument applied. However, one can always remove this by labelling the outcomes of the instruments by $b$, thereby introducing an additional index, and then considering the game with $\omega_{i,j,x,a,b}=\delta_{a,b}\omega_{i,j,x,a}$.

This approach to witnessing and quantifying instruments encompasses new classes of resources that existing methods cannot be directly applied to. For example, an interesting convex subset for single instruments on bipartite systems is given by those that are implementable through local operations and classical communication (LOCC).
Such instruments are of interest in, for example, the study of the resource theory of entanglement, in which they are free operations \cite{Plenio2007,Horodecki2009,Guhne2009}. For finitely many rounds of LOCC the set of instruments is compact. For unbounded numbers of rounds, one can consider the closure of these operations in order to fit it in our framework~\cite{Chitambar2014}.

As another example, a natural notion of compatibility for a set of instruments $\qty{I_{a|x}}_{a,x}$ is defined as the existence of a common instrument together with classical post-processings such that $I_{a|x}=\sum_\lambda p(a|x,\lambda)I_\lambda$~\cite{Heinosaari2014}. 
This definition is equivalent to unsteerability of channels~\cite{piani2015a}, i.e., the non-existence of an incoherent channel extension. Compatibility of sets of instruments clearly defines a convex set. Moreover, note that steering on the level of quantum states is a special case of channel steering, i.e., instruments with one-dimensional input systems.

\section{Higher-order dynamics}
\label{sec:supermaps}

Thus far we have focused on quantifying properties of quantum dynamics (e.g., channels and instruments).
Now we will see that the same game-theoretic approach can be generalized also to higher-order dynamics, i.e., transformations of dynamical objects.
Such higher-order dynamics have become an increasingly active field of research, from their role in studying quantum causality~\cite{oreshkov12} to their use as operations in resource theories~\cite{liu19}, but thus far have been given no operational resource theoretic study.

Formally, higher-order dynamics are ``supermaps'' that map a set of channels to another channel~\cite{chiribella08,chiribella08a}.
For simplicity, we focus here on supermaps of two channels, but the following generalizes immediately to any number of channels.
A supermap $\mathcal{S}$ thus transforms the channels $\Lambda_C,\Lambda_D$ to $\Lambda^{A\to B}=\mathcal{S}(\Lambda_C,\Lambda_D)$. 
For $\mathcal{S}$ to be valid, i) $\Lambda^{A\to B}$ must be a valid channel whenever $\Lambda_C,\Lambda_D$ are channels, and ii) $\mathcal{S}$ must give valid channels when applied locally to part of some bipartite channels, i.e., $\mathcal{I}\otimes\mathcal{S}$ (where $\mathcal{I}$ is the identity channel) must map the bipartite channels to channels~\cite{chiribella08a,araujo17,quintino18} (see Appendix~\ref{app:supermaps} for details).

The generalization of our approach to higher-order dynamics requires also a generalization of input-output games to collaborative games between several players.
As before, Alice and Bob prepare states from an ensemble $\mathcal{E}$ and perform the measurement $\textbf{M}$, respectively; Charlie and Dave measure quantum instruments $\textbf{I}^C=\{I^C_k\}_k$ and $\textbf{I}^D=\{I^D_\ell\}_\ell$ and for each tuple $(i,j,k,\ell)$ the parties get a score according to a reward function $\Omega=\qty{\omega_{ijk\ell}}_{ijk\ell}$, where $\omega_{ijk\ell}\in\mathbb{R}$.
 The tuple $\mathcal{G}=(\mathcal{E},\textbf{M},\textbf{I}^C,\textbf{I}^D,\Omega)$ thus defines a collaborative game in which the parties choose a supermap $\mathcal{S}$ in order to maximize the payoff function $P(\mathcal{S}, \mathcal{G})=\sum_{ijk\ell}p(i)\omega_{ijk\ell}\tr[\mathcal{S}(I^C_k,I^D_\ell)(\varrho_i) M_j]$.
As for input-output games we will assume that all collaborative games are in a canonical (positive, normalized) form.

As before, we can define the robustness of a supermap $\mathcal{S}$ with respect to a (convex, compact) free subset of supermaps $F$ as
\begin{equation}
\label{Eq:SRobustness}
\mathcal{R}_F(\mathcal{S}) = \min_{\tilde{S}}\qty{t\geq 0\bigg\vert \frac{\mathcal{S} + t\tilde{\mathcal{S}}}{1+t}\in F},
\end{equation}
where one minimizes over all supermaps $\tilde{\mathcal{S}}$.
Using similar techniques as earlier in the paper (see Appendix~\ref{app:supermaps} for details) one can show the following analog of Theorem~\ref{thm1} for supermaps:
\begin{theorem}\label{thmSupermap}
	Let $F$ be a convex and compact subset of supermaps.
	Then for every $\mathcal{S}\notin F$ there exists a collaborative game $\mathcal{G}$ such that, using $\mathcal{S}$, there is a strategy that outperforms any $\tilde{\mathcal{S}}\in F$.
	Moreover, this outperformance is quantified by the generalized robustness as
	\begin{equation}
	\label{Eq:result1W}
	        \sup_\mathcal{G}\frac{P(\mathcal{S}, \mathcal{G})}{\max_{\tilde{\mathcal{S}}\in F} P(\tilde{\mathcal{S}}, \mathcal{G})} = 1+\mathcal{R}_F(\mathcal{S}).
	\end{equation}
\end{theorem}

Let us note two important points regarding this result.
Firstly, Theorem~\ref{thmSupermap} is readily generalizable further to \emph{sets} of supermaps and, indeed, the proof in Appendix~\ref{app:supermaps} considers this case; we have avoided stating it in this form here simply to avoid further cluttering the notation, and because the examples discussed below do not make use of this.
As channels (and therefore states and measurements) are special cases of supermaps, this result thereby provides a unified game-theoretic characterization of these resources, placing them on the same footing.
Secondly, we emphasize that while game-theoretic quantification of certain channel-based resources has previously been considered, no such consideration has previously been given to higher-order dynamics.
Moreover, the approaches used for channel resources of reducing them to state-discrimination tasks via the Choi picture~\cite{Takagi2019a} does not---unlike input-output games---appear to be readily generalizable to higher-order operations.

\subsection{Applications to higher-order resources}

One of the key problems in the study of supermaps is to understand their structure: can they be understood as composing channels in parallel (with joint encoding and decoding maps), sequentially in a circuit~\cite{chiribella08} (with or without memory~\cite{giarmatzi18}), or do they even compose channels in a way that can be understood causally~\cite{araujo15,oreshkov12,oreshkov16,wechs18}?
Supermaps in some of these categories are known to provide advantages to those in others (e.g., sequential vs.\ parallel in metrology tasks~\cite{giovannetti06,duan07,dam07,bisio10}, or ``causally nonseparable'' ones in information theoretic tasks~\cite{chiribella12,chiribella13,araujo14,feix15,guerin16}).
In general, however, these advantages have not been understood in any unified fashion.
Theorem~\ref{thmSupermap} is applicable to all these sets of higher-order resources (which are, in fact, characterized through positive semidefinite constraints via the Choi picture~\cite{araujo15,chiribella08,wechs18}), and shows that they all indeed provide operational advantages in collaborative games.

Just as channels can be generalized to instruments, there is also a natural generalization from superchannels to superinstruments which already have found applications in analyzing several tasks~\cite{chiribella08a,chiribella09,quintino18,sedlak19}.
In Appendix~\ref{app:superinstruments} we show that Theorem~\ref{thmSupermap} can indeed be generalized to sets of such objects, emphasizing its utility in providing operational advantages to the most general form of dynamical resource.
It can thus, for example, be used to operationally witness the incompatibility of sequential superinstruments, often known as quantum testers~\cite{sedlak16}.

With interest in supermaps and superinstruments continuing to grow rapidly---they provide, e.g., the natural tool to describe free transformations between channel resources---we expect new higher-order resources to emerge and become relevant.
Our systematic approach to quantifying such resources should provide a key tool for understanding them as they are uncovered.

\section{Conclusions}

We have presented a general framework for finding task-oriented characterizations for quantum resources. Our results apply to a broad range of quantum objects with a convex (and compact) free set. The applicability of this framework is exemplified on the level of quantum channels, instruments and supermaps. As quantum measurements and states are special cases of channels, the technique applies to these objects as well, thereby providing a unified approach to characterizing such resources.

On top of giving quantum resources a task-oriented characterization, our framework comes with a simple quantifier. Namely, the outperformance of the resourceless objects by resourceful objects is exactly quantified by the generalized robustness measure.

In future research it will be interesting to see if the level of trust required (in the preparation of the objects used in the input-output games) can be reduced without impacting the generality of our results. One possible candidate for this would be to consider measurement-device-independence on the measurement performed on the output of the channel.

\begin{acknowledgments}
We are grateful to Ryuji Takagi and Bartosz Regula for pointing out that defining input-output games in their non-canonical form is too weak for the purposes of this manuscript. This work was supported by the DFG, the ERC (Consolidator Grant 683107/TempoQ), the Finnish Cultural Foundation, and the SNSF (Starting Grant DIAQ and NCCR SwissMAP).

\textit{Note added.---}%
After completing this manuscript we became aware of four related but independent works by J.\ Mori~\cite{Mori2019} and C.\ Carmeli et al.~\cite{Carmeli2019b} proving a connection between channel incompatibility and state discrimination, by X.\ Yuan et al.~\cite{Yuan2019} proving a connection between entanglement breaking channels and input-output games, and by P. Lipka-Bartosik et al. connection the robustness of instruments to teleportation-based quantum games \cite{lipka-bartosik19}.
\end{acknowledgments}

\bibliography{ChannelRef}

\appendix

\section{Conic programming and evaluating the robustness for sets of channels and instruments}
\label{appendix:conic}

\renewcommand{\theequation}{A\arabic{equation}}
\setcounter{equation}{0}

A \emph{convex cone} is a subset $C$ of a vector space $V$ if it is convex and one has ${ax\in C}$ for all ${x\in 
C}$ and $a\geq 0$. The \emph{dual cone} $C^*$ is defined as $C^*=\qty{y \mid 
\braket{x}{y}\geq 0~\forall x\in C}$. A generic cone program is of the following form~\cite{Cone_programming_Gaertner}
\begin{eqnarray}
\label{Eq:PrimalCone}
\max_X\, &&\, \tr[AX]  \\
\text{s.t.: }&& \Phi[X] \leq B,\quad X \in C, \notag
\end{eqnarray}
where $\Phi$ is a linear operator and $\geq$ denotes the partial order in the positive semidefinite cone. Following from Lagrange duality the dual cone program reads
\begin{eqnarray}
\label{Eq:DualCone}
\min_Y\, && \tr[BY] \\
\text{s. t.: }&& \Phi^{\dagger}[Y] - A\in C^*, \quad Y \geq 0. \notag
\end{eqnarray}
Similar to the case of SDPs, the solutions of the primal and dual problem coincide, i.e., strong duality holds, if and only if Slater's condition is satisfied and the primal problem is finite~\cite{Cone_programming_Gaertner}.

The channel robustness in Eq.~\eqref{Eq:ChannelRobustness} of the main text can be formulated as a cone program. Namely
\begin{eqnarray}
1+\mathcal{R}_F(\mathbf{\Lambda})  = \min_t &&\, 1+t \label{eq:coneConstr1}\\
\text{s.t.: }&& t\geq 0 \\
\label{Eq:ConeConstraint}
&& \frac{\mathbf{\Lambda} + t \tilde{\mathbf{\Lambda}}}{1+t}=\hat{\mathbf{\Lambda}}\in F \\
&& \tilde{\mathbf{\Lambda}}\text{ is a set of channels}.
\end{eqnarray}

Solving $\tilde{\mathbf{\Lambda}}$ from the constraint in Eq.~\eqref{Eq:ConeConstraint} of the above cone program gives $\tilde{\mathbf{\Lambda}}=\frac{1}{t}\qty[\mathbf{\Gamma}-\mathbf{\Lambda}]$, where $\mathbf{\Gamma}=(1+t)\hat{\mathbf{\Lambda}}$ and $\hat{\mathbf{\Lambda}}\in F$. Hence, the optimization problem in Eq.~\eqref{Eq:ChannelRobustness}, or more precisely the optimization problem plus one, can be cast in the Choi picture as
\begin{eqnarray}
1+\mathcal{R}_F(\mathbf{\Lambda})=\min_{J_{\mathbf{\Gamma}}}\, &&  \frac{1}{\abs{X}}\tr[J_{\mathbf{\Gamma}}] \label{eq:conedual} \\
\text{s.t.: }&& J_{\mathbf{\Gamma}}-J_{\mathbf{\Lambda}} \geq 0, \quad J_{\mathbf{\Gamma}}\in C_{J_F},\notag
\end{eqnarray}
where $J_{\mathbf{\Lambda}}=\oplus_x J_{\Lambda^{A\rightarrow B}_x}$, $J_{\mathbf{\Gamma}}=\oplus_x J_{\Gamma^{A\rightarrow B}_x}$, and   $C_{J_F}:=\{\alpha J_{\hat{\mathbf{\Lambda}}}\mid \alpha\geq 0, \hat{\mathbf{\Lambda}}\in F\}$ is the conic hull of $J_F$. This optimization problem is now in the form of the cone program~\eqref{Eq:PrimalCone}. The dual cone program can be obtained from Eq.~\eqref{Eq:DualCone} and the dual cone constraint can be further simplified (see Ref.~\cite{uola2019} for more details) such that the resulting dual program reads
\begin{eqnarray}
1+\mathcal{R}_F(\mathbf{\Lambda}) = \max_{Y}\, &&  \tr[Y J_{\mathbf{\Lambda}}] \\
\text{s.t.: }&& Y \geq 0, \quad \tr[YT]\leq 1\, \forall\ T\in J_{F}. \notag
\end{eqnarray}

For sets of instruments one follows the above calculations. The only difference is that each instrument element is treated as its own block.

The solutions of the primal and dual problems coincide if the so-called Slater's condition is fulfilled. In our scenario these conditions simply state that the positive semidefinite constraint in the primal problem can be satisfied in the strict form $J_{\mathbf{\Gamma}}-J_{\mathbf{\Lambda}} > 0$. In our examples this condition is satisfied as the maximally mixed state is in the free sets (in the Choi picture). Hence, one has a positive full rank point which can be scaled up to be strictly larger than a given $J_{\mathbf{\Lambda}}$.

\section{Quantum supermaps}
\label{app:supermaps}

\renewcommand{\theequation}{B\arabic{equation}}
\setcounter{equation}{0}

A quantum supermap is a linear higher-order transformation that maps a set of quantum channels $\mathbf{\Lambda}=\qty{\Lambda_1,\dots,\Lambda_n}$ (which, \emph{a priori}, may have different input and output Hilbert space dimensions $d_i^I,d_i^O$ so that $\Lambda_i : \mathcal{H}_i^I \to \mathcal{H}_i^O$) into a quantum channel $\mathcal{S}(\mathbf{\Lambda}): \mathcal{H}_0^I \to \mathcal{H}_0^O$~\cite{chiribella08,chiribella08a,quintino18}.
Moreover, just as a quantum channel must map quantum states to states even when applied to part of a bipartite state (which means they must be \emph{completely} positive maps), a quantum supermap must map channels to channels even when applied locally to part of some bipartite channels.

More formally, a linear map $\mathcal{S}$ must satisfying the following conditions to be a valid supermap~\cite{quintino18}:
\begin{itemize}
	\item TPP (trace-preserving preserving): If all $\Lambda_i\in \mathbf{\Lambda}$ are trace-preserving (TP) then $\mathcal{S}(\mathbf{\Lambda})$ must also be TP;
	\item CCPP (completely complete-positivity preserving): If the $\Lambda_i\in \mathbf{\Lambda}$ are bipartite completely positive maps from $\mathcal{H}_i^I\otimes {\mathcal{H}_i^I}' \to \mathcal{H}_i^O\otimes {\mathcal{H}_i^O}'$ then $\mathcal{S}\otimes\mathcal{I}(\mathbf{\Lambda})$ is a completely positive map, where $\mathcal{I}$ is the identity map on channels in the primed Hilbert spaces.
\end{itemize}
A supermap is thus a completely-CPTP preserving (CCPTP) map.

The characterization of supermaps is more easily expressed in the Choi picture.
There, supermaps are represented as process matrices~\cite{araujo17}, which were first introduced as maps from CP maps to probabilities in the study of indefinite causal orders~\cite{araujo15,oreshkov12}.
The process matrix $W$ of a supermap $\mathcal{S}$ is a matrix satisfying the following constraints:
\begin{itemize}
	\item PSD (positive semidefiniteness): $W \ge 0$;
	\item Normalization: $\tr [W] = 1$ (note that, to ensure correspondence with the case of channels, we use a different normalization than is used elsewhere in the literature on process matrices).
	\item Validity: $L_V(W)=W$, where $L_V$ is the projector onto the linear subspace of valid process matrices as defined in Refs.~\cite{araujo15,araujo17}.
\end{itemize}
For the case of supermaps on two channels that we consider in the main text, the validity constraint takes the form
\begin{align}\label{eq:Wvalidity}
	 \tr_{I_1O_1 I_2 O_2 O_0}W =& \id^{I_0}\notag\\
	 \tr_{I_2 O_2 O_0}W =& \tr_{O_1}(\tr_{I_2 O_2 O_0}W)\otimes \id^{O_1}\notag\\
	\tr_{I_1 O_1 O_0}W =& \tr_{O_2}(\tr_{I_1 O_1 O_0}W)\otimes \id^{O_2}\notag\\
	\tr_{O_0} W =& \tr_{O_1}(\tr_{O_0} W)\otimes \id^{O_1}\notag\\
	 &+ \tr_{O_2}(\tr_{O_0} W)\otimes \id^{O_2}\notag \\
	  &- \tr_{O_1 O_2}(\tr_{O_0} W)\otimes \id^{O_1 O_2},
\end{align}
where the labels $I_i$ and $O_i$ represent the Hilbert spaces $\mathcal{H}_i^I$ and $\mathcal{H}_i^O$, respectively.

\medskip

The conic programming approach described in the main text and in more detail in the previous section for channels can be generalized simply to sets of supermaps.
To this end, it will again be useful to work in the Choi picture where there supermaps are represented by process matrices as described above.
Then, it is readily seen that Eqs.~\eqref{eq:coneConstr1} to~\eqref{eq:conedual} hold with $\Lambda$ replaced by $\mathbf{S}=\qty{\mathcal{S}_x}_x$ a set of supermaps, $J_\Lambda$ by $W=\oplus_x W_{\mathcal{S}_x}$, etc.
One thus finds (subject, as before, to Slater's condition holding for $F$)
\begin{align}\label{eq:robustnessW}
	1+\mathcal{R}_F(\mathbf{S}) = & \max_{Y}\, \sum_x\tr[Y_x W_{\mathcal{S}_x}] \\
	&\text{s.t.: } Y \geq 0, \quad \tr[YT]\leq 1\, \forall\ T\in W_F, \notag
\end{align}
where $W_F$ is the collection of sets of process matrices representing the sets of supermaps in $F$.
The case treated in the main text corresponds to $|X|=1$.

With the definition of the Choi map used in the main text and the process matrix normalization constraint, the probability of observing outcomes $i_1,\dots,i_n,i_0$ when performing instruments $\mathbf{I}_1,\dots,\mathbf{I}_n$ followed by a final measurement $\mathbf{M}=\{M_{i_0}\}_{i_0}$ when the input is $\rho$ (i.e., measuring $M$ on $\mathcal{S}(I_{i_1},\dots,I_{i_n})(\rho)$) is given by
\begin{equation}
	\frac{1}{D}\tr[W_{\mathcal{S}} (\rho^T\otimes J_{I_{i_1}}^T\otimes\cdots\otimes J_{I_{i_n}}^T\otimes M_{i_0})],
\end{equation}
where $D=d_0^I \prod_i d_i^I d_i^O$.
Writing the witness $Y=\oplus_x Y_x$ with~\cite{araujo15,branciard16a}
\begin{align}
	Y_x = D &\sum_{i,j_1,\dots,j_n,k} p(i,x)\omega_{i,j_1,\dots,j_n,k,x}\notag\\
	&\times\rho^T_{i|x} \otimes J^T_{I_{i_1|x}} \otimes \cdots \otimes J^T_{I_{i_n|x}} \otimes \eta^T_{k|x},
\end{align}
where the $\omega_{i,j_1,\dots,j_n,k}$ chosen to ensure the $J_{i_\ell}$ are all Choi maps of instruments and the $\eta_k$ POVM elements, one then arrives at
\begin{equation}
\label{Eq:result1Wb}
        \sup_\mathcal{G}\frac{P(\mathbf{S}, \mathcal{G})}{\max_{\tilde{\mathbf{S}}\in F} P(\tilde{\mathbf{S}}, \mathcal{G})} = 1+\mathcal{R}_F(\mathbf{S}),
\end{equation}
where the supremum is over all collaborative games $\mathcal{G}$.
This thereby proves Theorem~\ref{thmSupermap} of the main text.

In the examples we mention in the main text, Slater's condition is easily seen to be satisfied by taking the maximally noisy process (whose process matrix is proportional to $\id$) which is contained in the free sets we consider~\cite{araujo15,wechs18}.

\section{Quantum superinstruments}
\label{app:superinstruments}

As mentioned in the main text, Theorem~\ref{thmSupermap} can be generalized readily to sets of ``quantum superinstruments'' (also called probabilistic supermaps).
Formally, a quantum superinstrument is a collection $\mathbf{T}=\qty{\mathcal{T}_a}_{a}$ of maps, where each $\mathcal{T}_a$ is CCPP and $\sum_a \mathcal{T}_a$ is TPP and thus a valid quantum supermap~\cite{chiribella08a,chiribella09}.
In the process matrix picture, each $\mathcal{T}_{a}$ is simply represented by a positive semidefinite matrix $W_{\mathcal{T}_a}$ with $\sum_a W_{\mathcal{T}_a}$ a valid process matrix~\cite{quintino18}.

For clarity and simplicity of presentation, let us here present just the case of superinstruments acting on two CP maps with input and output dimensions $d$; the general case follows immediately as elucidated in the previous section on quantum sumpermaps.
In analogy to the generalization from quantum channels to instruments, to go to the case of superinstruments one writes the payoff function for the collaborative game (where the sets of ensembles, instruments and states making up the game are now indexed by $x$ as well) as
\begin{align}
	P(\textbf{T},\mathcal{G})=&\sum_{i,j,k,\ell,x,a}  p(i,x,a)\,\omega_{ijk\ell xa} \notag\\
	&\times\tr[\mathcal{T}_{a|x}(I^C_{k|x,a},I^D_{\ell|x,a})(\varrho_{i|x,a}) M_{j|x,a}].
\end{align}
Considering a free set $F$ of collections of quantum superinstruments, one then defines  the robustness with respect to $F$ analogously as to in the previous cases, writes the witness $Y=\oplus_{a,x}Y_{a|x}$ and decomposes each $Y_{a|x}$ as $Y_{a|x}=d^5\sum_{ijk\ell ax}p(i,x,a)\omega_{ijk\ell ax}\rho^T_{i|x,a}\otimes J^T_{I_{k|x,a}}\otimes J^T_{I_{\ell|x,a}} \otimes \eta_{j|x,a}$.

In the case where $\sum_a \mathcal{T}_a$ has a sequential realization, superinstruments are often called quantum testers~\cite{chiribella12} or process POVMs~\cite{ziman08}.
Sequential superinstruments are known to provide advantages over parallel ones in some tasks~\cite{chiribella09,sedlak19}; in some of these, such as the problem of probabilistically inverting unknown unitaries~\cite{quintino18}, general superinstruments provide yet a further advantage.

In addition to quantifying the advantage of superinstruments with particular causal structures using collaborative games, one can also use these games to study, e.g., the compatibility of sequential superinstruments~\cite{sedlak16}.
A set of sequential superinsturment $\qty{\mathcal{T}_{a|x}}_{a,x}$ is called compatible if (in analogy to compatibility for POVMs) there exists a joint sequential superinstrument $\qty{\mathcal{K}_{\lambda}}_\lambda$ such that $\mathcal{T}_{a|x}=\sum_\lambda p(a|x,\lambda)\mathcal{K}_\lambda$. Compatible superinstruments form a free set that our approach can readily be applied to.

The study of superinstruments is still in its infancy---e.g., the concept of compatibility has not yet been studied for general superinstruments---but our results show already that, as such properties become understood, they can be quantified using the game theoretic approach we introduce.

\end{document}